# Deep Learning-based Multi-Organ CT Segmentation with Adversarial Data Augmentation

Shaoyan Pan[1,2], Shao-Yuan Lo[3], Min Huang[2], Chaoqiong Ma[1], Jacob Wynne[1], Tonghe Wang[4], Tian Liu[1] and Xiaofeng Yang[1,2]

[1]Department of Radiation Oncology, Emory University, Atlanta, GA 30322, USA
[2]Department of Biomedical Informatics, Emory University, Atlanta, GA 30322, USA
[3]Department of Electrical and Computer Engineering, Johns Hopkins University, Baltimore, MD 21218, USA
[4]Department of Medical Physics, Memorial Sloan Kettering Cancer Center, New York, NY, USA

## Abstract

In this work, we propose an adversarial attack-based data augmentation method to improve the deep-learning-based segmentation algorithm for the delineation of Organs-At-Risk (OAR) in abdominal Computed Tomography (CT) to facilitate radiation therapy. We introduce Adversarial Feature Attack for Medical Image (AFA-MI) augmentation, which forces the segmentation network to learn out-of-distribution statistics and improve generalization and robustness to noises. AFA-MI augmentation consists of three steps: 1) generate adversarial noises by Fast Gradient Sign Method (FGSM) on the intermediate features of the segmentation network's encoder; 2) inject the generated adversarial noises into the network, intentionally compromising performance; 3) optimize the network with both clean and adversarial features. The effectiveness of the AFA-MI augmentation was validated on nnUnet. Experiments are conducted segmenting the heart, left and right kidney, liver, left and right lung, spinal cord, and stomach in an institutional dataset collected from 60 patients. We firstly evaluate the AFA-MI augmentation using nnUnet and Token-based Transformer Vnet (TT-Vnet) on the test data from a public abdominal dataset and an institutional dataset. In addition, we validate how AFA-MI affects the networks' robustness to the noisy data by evaluating the networks with added Gaussian noises of varying magnitudes to the institutional dataset. Network performance is quantitatively evaluated using Dice Similarity Coefficient (DSC) for volume-based accuracy. Also, Hausdorff Distance (HD) is applied for surface-based accuracy. On the public dataset, nnUnet with AFA-MI achieves DSC = 0.85 and HD = 6.16 millimeters (mm); and TT-Vnet achieves DSC = 0.86 and HD = 5.62 mm. AFA-MI is observed to improve the segmentation DSC score ranging from 0.055 to 0.003 across all organs relative to clean inputs. AFA-MI augmentation further improves all contour accuracies up to 0.217 as measured by the DSC score when tested on images with Gaussian noises. AFA-MI augmentation is therefore demonstrated to improve segmentation performance and robustness in CT multi-organ segmentation.

## I. Introduction

The radiotherapy treatment planning requires accurate OAR delineation aiming to control the OAR exposure to the radiation when tailoring the prescribed dose to the target. Treatment outcomes are therefore heavily reliant upon the accuracy of target and organ contouring. While physician-led manual segmentation achieves accurate organ contours, it is time-consuming, labor-intensive, and observer-dependent. Accordingly, automatic segmentation approaches are proposed as an effective and efficient solution for accurate and reproducible organ delineation. Currently, machine learning and deep learning models are prevalent in various medical tasks, including image classification and detection [1-5], registration [6] and synthesis [7, 8], and demonstrates state-of-the-art accuracy and efficiency. Similarly, for the segmentation, deep learning algorithms are frequently applied in this setting due to their superior ability to detect organ contours [9, 10]. More specifically, U-shaped convolutional neural networks (CNNs) such as Unet [11], Vnet [12], and recent vision transformer-based networks such as convolutional vision transformer [13-15], Token-based Transformer V-net (TT-Vnet) [16], or Multi-linear Perceptron Mixers [17-19] are among the most popular automatic segmentation algorithms that are capable of state-of-the-art performance across many segmentation tasks.

However, U-shaped networks require abundant data with high variability in organ size, appearance, and position during training to accurately model abdominal structures and reach optimal performance. Deep learning networks are also vulnerable to data corruption regularly encountered in real-world applications (e.g., noise, artifacts): network performance degrades when applied to real images harboring these defects. In this work, we propose a novel Adversarial Feature Attack for Medical Image (AFA-MI) augmentation to improve segmentation performance and robustness. We posit that an adversarial attack of noises injected during training into intermediate features of segmentation networks will not only result in accurate segmentation maps but will further improve generalization to new data and improve robustness to real-world data corruptions [20].

## II. Method

As illustrated in Figure 1, a U-shaped CNN is trained to map the input abdominal scans to multi-organ segmentation maps. Typically, the network consists of an encoder and a decoder. The encoder is a contracting path consisting of multiple down-sampling blocks to capture the compressed semantic information from the input scans. The decoder is an expanding path to decompress the features back to the original size of the input and obtain a final segmentation map for each organ. Given the input abdomen scan as $x$, network as $f$, and the target segmentation map $g$, the network is optimized by the following objective function:

$$\operatorname*{argmin}_{f} L_{seg}(x, f, g) = \operatorname*{argmin}_{f} [\gamma Dice(f(x), g) + (1-\gamma)CE(f(x), g)]$$

where $\gamma$ is a weight that empirically chosen as 0.5, $Dice$ is the dice loss, and $CE$ is the cross-entropy loss.

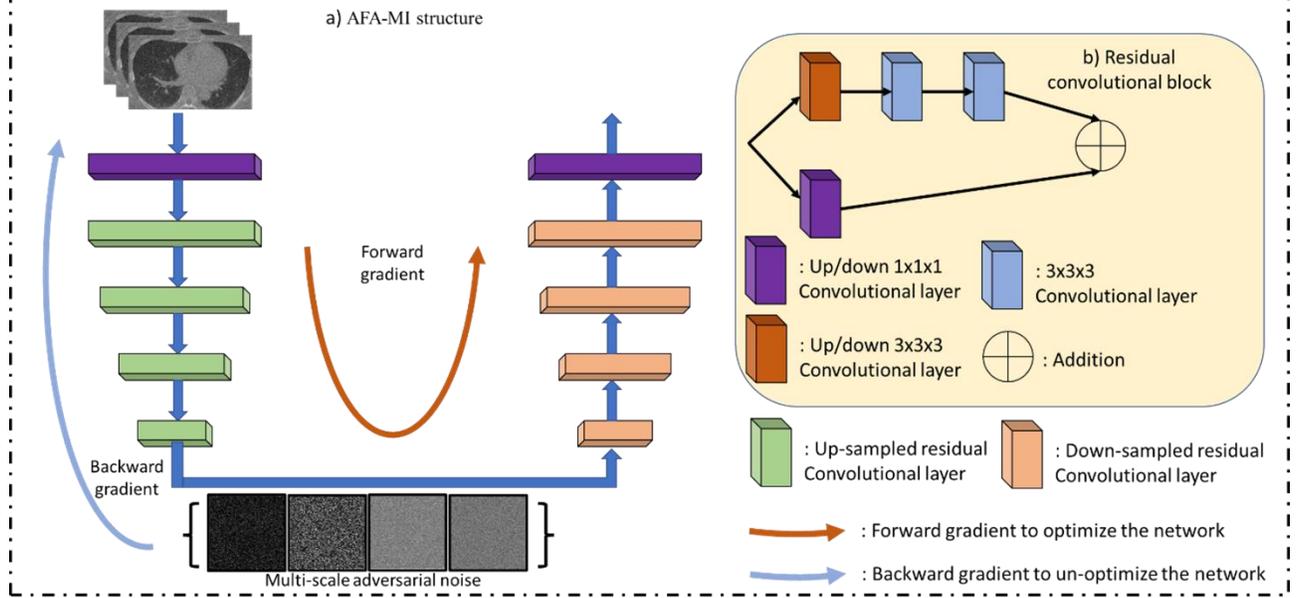

**Figure 1**: The pipeline of the proposed AFA-MI augmentation. During the training process, the network will first generate multi-scale adversarial attack noises targeting the i'th layer, intended to negatively impact final segmentation accuracy. The noises are then fused with the clean features of the same layer. As a result, the network is trained in an end-to-end fashion to obtain segmentation maps referencing not only the clean features, but also the features with the superimposed noises.

### II.A Adversarial attack-based data augmentation

Adversarial attack-based data augmentation, which trains a network with adversarial noises in the input or modifies intermediate features [21-24], can be a reliable approach to increasing a network's robustness to high-frequency noise and texture distortion. In addition, it can improve network performance on clean images. In our work, the AFA-MI includes three steps: 1) generate adversarial noises on the intermediate features of the segmentation network's encoder; 2) inject the generated adversarial noises into the network, intentionally compromising performance; 3) optimize the network with both clean and adversarial features.

### II.A.1 Adversarial feature noise generation

We attack the $i'th$ down-sampling block of the encoder using the well-known Fast Gradient Sign Method (FGSM) [25]. FGSM generates adversarial noises, through the backward gradient (Fig. 1), which maximizes the segmentation objective function thereby reducing network performance. We denote the set of the first layer to the $i'th$ layer as $f_i$, and the rest of the network as $f_{1-i}$, so $f(x) = f_{1-i}(f_i(x))$:

$$f_i^{adv, S_r}(x) = S_r(f_i(x) + \lambda + \epsilon * sign(\nabla_{f_i} L_{seg}(x, f, g)))$$

where $\lambda$ is noise initialized by a Gaussian distribution with mean 0 and standard deviation 1, $\epsilon$ is the adversarial noise strength, $S$ is the constraint of mean square intensity of the noise and $r$ is the constraint ratio, i.e., $S_{0.1}(.) = [0.9 \min(f_i(x)), 1.1 \max(f_i(x))]$. Empirically, $\epsilon$ is chosen as 0.003, and $r$ is 0.1.

### II.A.2 Adversarial feature noise injection
We compute the statistics of the generated adversarial noises to inject the adversarial features into clean features [21], intentionally compromising network performance. The adversarial feature noise injection is described as follows:

$$f_i^{noisy,S_r}(x) = \sigma^{adv} \frac{f_i^{clean} - \mu^{clean}}{\sigma^{clean}} + \mu^{adv}$$

where $\mu$ is the first-order moment, $\sigma$ is the second-order moment, $(\mu^{clean}, \sigma^{clean})$ is computed from $f_i^{clean}$, and $(\mu^{adv}, \sigma^{adv})$ is computed from $f_i^{adv,S_r}$. The noisy adversarial feature $f_i^{noisy,S_r}$ is then passed into the network to replace the clean feature $f_i^{clean}$.

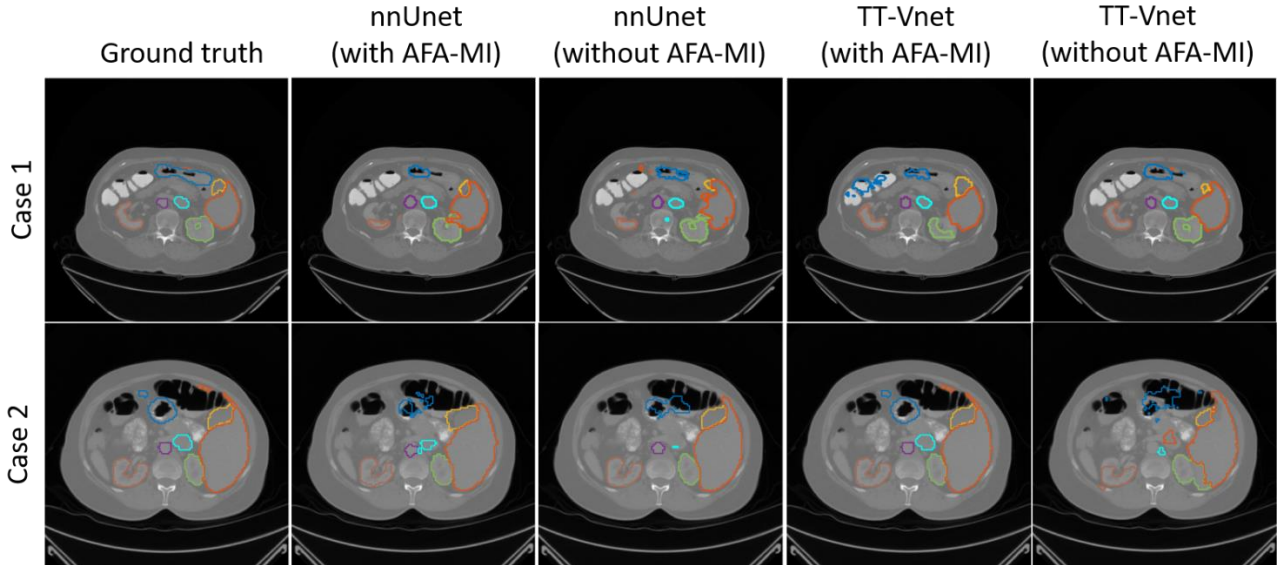

**Figure 2**: Example segmentation results against varying levels of additive Gaussian noises from the public dataset. Two patient scans with the manual contours, the automatic segmentations from nnUnet [28] and TT-Vnet [16] with/without the AFA-MI augmentation (in column-wise). The segmentations contain the left kidney (orange), right kidney (green), gallbladder (yellow), liver (light blue), stomach (blue), aorta (purple), and inferior vena cave (cyan).

### II.A.3 Overall optimization objective
After generating the adversarial features, the network, instead of training only with clean features, is trained with the clean and adversarial features together to obtain accurate segmentation maps for the intended organs. We train with multi-scale feature attacks to further bolster model generalization and robustness. The final objective function of the proposed AFA-MI augmentation training scheme is:

$$\sum_{k=0.1, 0.05, 0.025, 0.0125} \underset{f}{\mathrm{argmin}}\, L_{seg}\left(x, f_i^{noisy,S_k}{}_i, f_{1-i}, g\right)$$

## III. Data Acquisition and Preprocessing
We aim to segment the left and right kidney, gallbladder, liver, stomach, aorta, and inferior vena cava on 3D patient scans from a public dataset from the Beyond the Cranial Vault (BCTV) [26] segmentation challenge presented in 2015 at the 18th International Conference on Medical Image Computing and Computer-Assisted Intervention (MICCAI). In addition, we aim to segment heart, left and right kidney, liver, left and right lung, spinal, and stomach from an institutional CT dataset collected from 60 patients. Each CT volume and the corresponding organ contours are resampled to 2×2×3 millimeters (mm). The first 80% of the data are used for training, and the rest 20% of the data are used for testing. During training, we randomly select 4 patches with a size of 64×64×32 mm from each scan in each iteration. During testing, the segmentation map is predicted in a sliding-window manner with a window size equal to the patch size with 80% overlap. Mixup [27] data augmentation with Mixup parameter = 0.2 is applied to improve generalizability. For both training and testing, the intensity range of all images is normalized to -1 to 1 across the dataset. During testing, the network outputs are resampled to the original size of the inputs. Finally, Softmax and argmax functions are applied to the outputs, yielding final segmentation results.

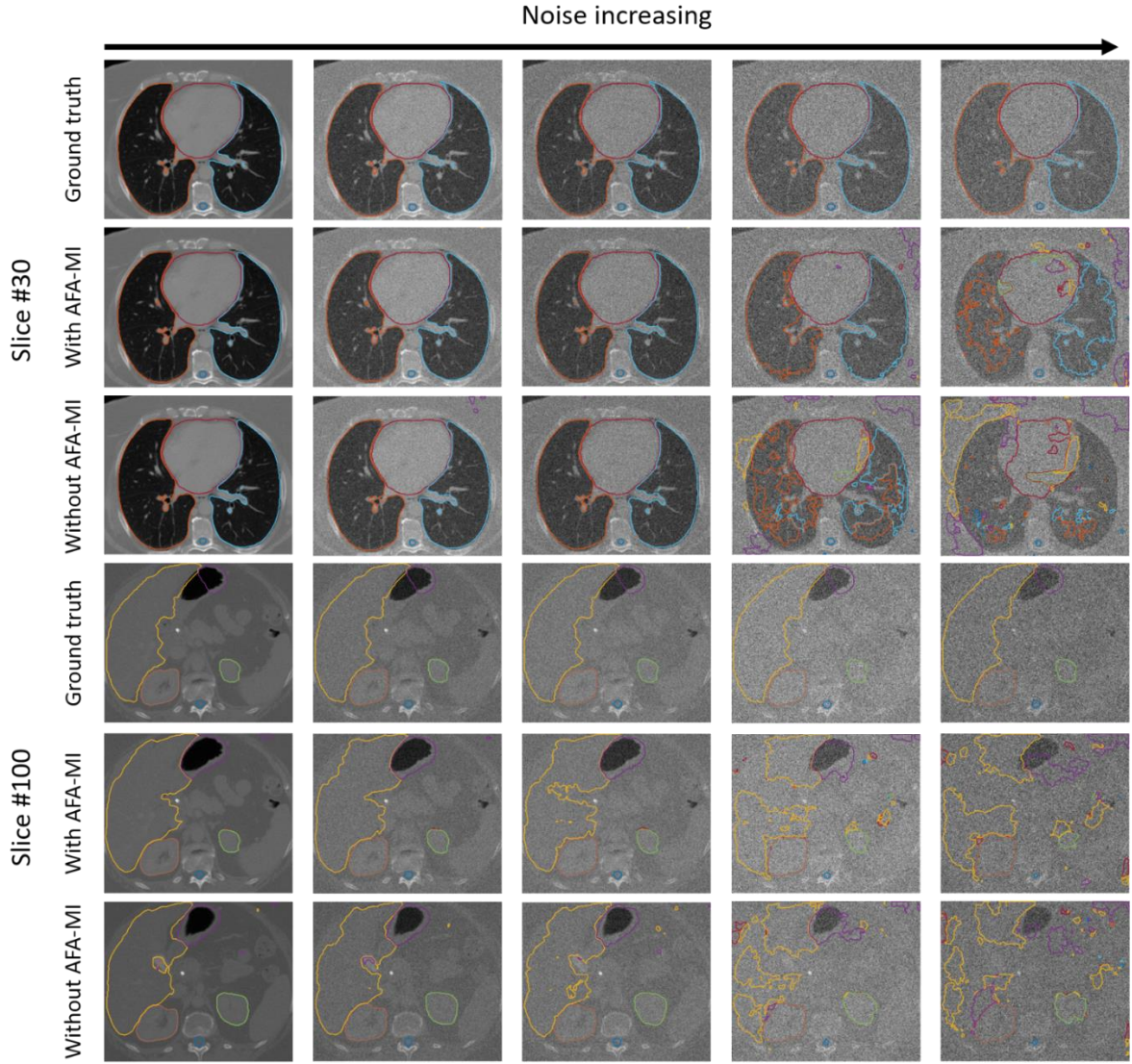

**Figure 3**: Example segmentation results against varying levels of additive Gaussian noises from the institutional dataset. In the top half, the central slices contain the heart (red), lung (orange and light blue), and spinal cord (blue). In the bottom half, the central slices contain the kidneys (light orange and green), liver (yellow), spinal cord (blue), and stomach (purple). From left to right, the input scan has additive Gaussian noises with a mean of 0 and standard deviations of 0.01, 0.005, 0.001, and 0.0005, respectively. From top to bottom within each slice: manual contours (ground truth), with AFA-MI augmentation, and nnUnet without augmentation.

## IV. Implementation Details and Performance Evaluation

We evaluate the network performance using the Dice Similarity Coefficient (DSC), where a high DSC score indicates good performance. In addition, we applied Hausdorff Distance (HD) for surface-based accuracy, where a lower HD indicates better performance. We apply the proposed AFA-MI augmentation to nnUnet [28] and Token-based transformer Vnet [16], two state-of-the-art medical image segmentation networks. In our experiments, we validate two effects of AFA-MI. First, we evaluate its augmentation effect on clean CT scans (i.e., without additive noise). On this clean test set, we compare the performance of both networks with AFA-MI against a corresponding network without AFA-MI. Second, we evaluate AFA-MI's augmentation effect on noisy scans from the institutional dataset. We add multi-scale Gaussian noises, with a mean of 0 and standard deviations of 0.01, 0.005, 0.001, and 0.0005, to the test set. We then compare the performance of the networks with and without AFA-MI augmentation. The Mann-Whitney U-test is applied to evaluate whether the proposed AFA-MI yields significant improvements over the network without augmentation.

## V. Results

The segmentation visualization of nnUnet and TT-Vnet is shown in Figure 2. The quantitative evaluation is given in Table 1. The visual comparison of nnUnet with and without AFA-MI augmentation against noisy scans is shown in Figure 3. Furthermore, the quantitative evaluation of the network on the noisy images is shown in Table. 2.

On the BCTV dataset, nnUnet and TT-Vnet with AFA-MI augmentation demonstrate better DSC on the left and right kidney, gallbladder, liver, aorta, and inferior vena cava. In terms of statistical tests, nnUnet with AFA-MI augmentation shows statistically significant improvement (p-values $< 0.05$) on the gallbladder, liver, and aorta, while TT-Vnet with AFA-MI shows statistical improvement on stomach. On the other hand, in terms of the HD, AFA-MI augmentation improves both nnUnet's and TT-Vnet's performance on the left and right kidney, gallbladder, liver, aorta, and inferior vena cava. The statistical evaluation also demonstrates that AFA-MI augmentation not only improves a traditional nnUnet performance on the gallbladder, liver, and aorta, but also improves TT-Vnet on the gallbladder and liver.

In the experiment on the noisy institutional images, with clean images (i.e., without Gaussian noises), the network with AFA-MI augmentation achieves a better DSC score than the network without augmentation. The quantitative improvement ranges from 0.055 to 0.003 across all organs. AFA-MI demonstrates statistically significant improvement in segmentations of the heart, left and right kidneys, left and right lungs, and spinal cord. With increased additive Gaussian noises, AFA-MI quantitatively improves the accuracy by a more considerable margin in most organs. In the most severe case, AFA-MI achieves a quantitative improvement from 0.217 to 0.045 and statistically significant improvements across all the organs.

**Table 1.** The DSC and HD comparison between nnUnet/TT-Vnet with and without AFA-MI augmentation. Improvement indicates the difference in accuracy between a network with and without AFA-MI, where positive values indicate superior performance with AFA-MI. p <0.05 indicates AFA-MI brings statistically significant improvement to the network. Better results are bolded.

|  | DSC score | Left kidney | Right kidney | Gallbladder | Liver | Stomach | Aorta | Vena cava |
|---|---|---|---|---|---|---|---|---|
| nnUnet | AFA-MI | **0.82±0.11** | **0.84±0.22** | **0.82±0.12** | **0.96±0.04** | 0.84±0.32 | **0.91±0.02** | **0.75±0.15** |
|  | No AFA-MI | 0.81±0.23 | **0.84±0.24** | 0.79±0.19 | 0.94±0.01 | **0.85±0.17** | 0.89±0.07 | 0.73±0.12 |
|  | P-value | 0.512 | 0.467 | 0.013 | 0.032 | N/A | 0.067 | <0.01 |
|  | HD (mm) | Left kidney | Right kidney | Gallbladder | Liver | Stomach | Aorta | Vena cava |
| nnUnet | AFA-MI | **6.56±5.42** | **6.43±6.32** | **6.43±3.45** | **3.78±2.98** | 4.74±7.43 | **6.15±3.12** | **9.00±4.56** |
|  | No AFA-MI | 6.80±6.67 | 6.65±10.68 | 8.04±3.48 | 5.10±2.29 | **4.66±9.09** | 6.61±3.70 | 9.49±9.73 |
|  | P-value | 0.54 | 0.311 | <0.01 | <0.01 | N/A | <0.01 | 0.078 |
|  | DSC score | Left kidney | Right kidney | Gallbladder | Liver | Stomach | Aorta | Vena cava |
| TT-Vnet | AFA-MI | **0.83±0.15** | **0.85±0.11** | **0.83±0.09** | **0.96±0.03** | **0.89±0.19** | **0.90±0.10** | **0.78±0.07** |
|  | No AFA-MI | 0.83±0.13 | 0.84±0.15 | 0.81±0.33 | 0.95±0.04 | 0.87±0.21 | 0.89±0.11 | 0.76±0.08 |
|  | P-value | 0.891 | 0.365 | 0.052 | 0.143 | 0.041 | 0.241 | 0.093 |
|  | HD (mm) | Left kidney | Right kidney | Gallbladder | Liver | Stomach | Aorta | Vena cava |
| TT-Vnet | AFA-MI | **5.36±5.13** | **6.12±6.31** | **7.12±4.23** | **3.21±3.68** | **4.51±6.44** | **5.44±2.43** | 7.81±7.32 |
|  | No AFA-MI | 5.89±4.56 | 6.39±7.12 | 7.55±4.12 | 4.21±2.01 | 4.85±6.17 | 5.78±5.44 | **7.67±8.31** |
|  | P-value | 0.112 | 0.211 | 0.031 | <0.01 | 0.321 | 0.087 | N/A |

**Table 2.** The DSC score comparison between nnUnet with and without AFA-MI augmentation. Improvement indicates the difference in accuracy between nnUnet with and without AFA-MI, where positive values indicate superior performance with AFA-MI. $p<0.05$ indicates AFA-MI brings statistically significant improvement to the network. Noise $\sim N(mean, std)$ indicates the mean and standard deviation of the additive Gaussian noises introduced into the input scans. Better results are bolded.

| | DSC score | Heart | Left kidney | Right kidney | Liver | Left lung | Right lung | Spinal cord | Stomach | Average |
|---|---|---|---|---|---|---|---|---|---|---|
| Noise $N(0,0)$ | nnUnet (AFA-MI) | **0.92±0.04** | **0.95±0.03** | **0.93±0.09** | **0.92±0.04** | **0.98±0.01** | **0.98±0.01** | **0.84±0.06** | **0.80±0.09** | **0.92** |
| | nnUnet (No AFA-MI) | **0.92±0.04** | 0.94±0.04 | 0.91±0.10 | **0.92±0.04** | **0.98±0.01** | **0.98±0.01** | 0.79±0.07 | 0.77±0.13 | 0.90 |
| | Improvement | 0.005 | 0.010 | 0.020 | 0.006 | 0.005 | 0.003 | 0.055 | 0.030 | 0.02 |
| | P-value | 0.041 | 0.033 | 0.026 | 0.248 | 0.003 | 0.013 | 0.008 | 0.424 | N/A |
| Noise $N(0,0.0005)$ | nnUnet (AFA-MI) | **0.92±0.04** | **0.94±0.04** | **0.92±0.09** | **0.92±0.04** | **0.98±0.01** | **0.98±0.01** | **0.83±0.06** | **0.77±0.09** | **0.91** |
| | nnUnet (No AFA-MI) | 0.91±0.04 | 0.93±0.05 | 0.90±0.10 | 0.91±0.04 | 0.97±0.02 | 0.97±0.01 | 0.78±0.07 | 0.74±0.13 | 0.89 |
| | Improvement | 0.008 | 0.014 | 0.024 | 0.010 | 0.010 | 0.005 | 0.049 | 0.032 | 0.02 |
| | P-value | 0.033 | 0.026 | 0.026 | 0.026 | 0.010 | 0.328 | 0.006 | 0.213 | N/A |
| Noise $N(0,0.001)$ | nnUnet (AFA-MI) | **0.91±0.04** | **0.93±0.04** | **0.91±0.09** | **0.91±0.04** | **0.97±0.01** | **0.97±0.01** | **0.83±0.07** | **0.76±0.10** | **0.90** |
| | nnUnet (No AFA-MI) | 0.90±0.04 | 0.92±0.05 | 0.88±0.10 | 0.89±0.05 | 0.95±0.04 | 0.96±0.02 | 0.77±0.07 | 0.72±0.14 | 0.88 |
| | Improvement | 0.010 | 0.009 | 0.024 | 0.017 | 0.022 | 0.010 | 0.051 | 0.033 | 0.02 |
| | P-value | 0.010 | 0.110 | 0.010 | 0.006 | 0.050 | 0.213 | 0.008 | 0.213 | N/A |
| Noise $N(0,0.005)$ | nnUnet (AFA-MI) | **0.85±0.06** | **0.86±0.08** | **0.84±0.1** | **0.86±0.07** | **0.9±0.03** | **0.85±0.07** | **0.75±0.09** | **0.64±0.13** | **0.82** |
| | nnUnet (No AFA-MI) | 0.79±0.08 | 0.85±0.09 | 0.8±0.11 | 0.82±0.07 | 0.66±0.09 | 0.59±0.15 | 0.61±0.12 | 0.58±0.18 | 0.71 |
| | Improvement | 0.063 | 0.015 | 0.049 | 0.037 | 0.242 | 0.260 | 0.140 | 0.060 | 0.11 |
| | P-value | 0.013 | 0.328 | 0.004 | 0.008 | 0.003 | 0.003 | 0.003 | 0.131 | N/A |
| Noise $N(0,0.01)$ | nnUnet (AFA-MI) | **0.74±0.11** | **0.78±0.11** | **0.75±0.13** | **0.77±0.10** | **0.71±0.10** | **0.42±0.15** | **0.66±0.11** | **0.53±0.17** | **0.67** |
| | nnUnet (No AFA-MI) | 0.60±0.11 | 0.71±0.18 | 0.66±0.16 | 0.73±0.10 | 0.19±0.10 | 0.12±0.07 | 0.44±0.14 | 0.44±0.19 | 0.48 |
| | Improvement | 0.148 | 0.073 | 0.092 | 0.045 | 0.527 | 0.304 | 0.217 | 0.094 | 0.19 |
| | P-value | 0.003 | 0.016 | 0.010 | 0.010 | 0.003 | 0.003 | 0.003 | 0.033 | N/A |

## VI. Conclusion

This work presents an efficient data augmentation routine by adversarial attacks, which improves the segmentation performance and robustness of a U-shaped deep learning network in abdominal CT scans. The proposed adversarial feature attack (AFA-MI) data augmentation consists of two steps: 1) the AFA-MI generates noise to "attack" the features learned by maximizing the loss between the network's prediction and the target contour; 2) the network needs to generate accurate segmentation even with the attack to increase its robustness. With the proposed adversarial attack-based data augmentation AFA-MI, segmentation networks achieve better accuracy on a public abdomen CT dataset and an institutional CT dataset. Moreover, the networks with AFA-MI achieve significantly better results on noisy scans compared to those without AFA-MI, demonstrating that networks with AFA-MI can be more robust against noisy samples. In future work, we plan to apply AFA-MI to segmentation tasks on the dataset with potential severe noise or artifacts (e.g., Cone beam CT dataset). In addition, other networks will be experimented with to validate whether the AFA-MI augmentation can be generalized to other deep learning segmentation networks. Furthermore, we plan to use additional evaluation metrics, e.g., the mean surface distance for surface-based accuracy, to characterize the effectiveness of AFA-MI better.

## ACKNOWLEDGEMENTS

This research is supported in part by the National Institutes of Health under Award Number R01CA215718, R56EB033332 and R01EB032680.

# References

[1] Y. Li, J. S. Hsu, N. Bari *et al.*, "Interpretable Evaluation of Diabetic Retinopathy Grade Regarding Eye Color Fundus Images," IEEE International Conference on Bioinformatics and Bioengineering, (2022).

[2] M. Hu, J. Amason, T. Lee *et al.*, "Deep learning approach for automated detection of retinal pathology from ultra-widefield retinal images," Investigative Ophthalmology & Visual Science, 62(8), 2129-2129, (2021).

[3] C.-W. Chang, S. Zhou, Y. Gao *et al.*, "Validation of a deep learning-based material estimation model for Monte Carlo dose calculation in proton therapy," Physics in Medicine & Biology, 67(21), 215004, (2022).

[4] C.-W. Chang, Y. Gao, T. Wang *et al.*, "Dual-energy CT based mass density and relative stopping power estimation for proton therapy using physics-informed deep learning," Physics in Medicine & Biology, 67(11), 115010, (2022).

[5] T. Li, M. Hu, and L. Zhang, "Using the SVM Method for Lung Adenocarcinoma Prognosis Based on Expression Level," International Conference on Computational Biology and Bioinformatics, (2018).

[6] Y. Fu, Y. Lei, T. Wang *et al.*, "Deep learning in medical image registration: a review," Physics in Medicine & Biology, 65(20), 20TR01, (2020).

[7] T. Wang, Y. Lei, Y. Fu *et al.*, "A review on medical imaging synthesis using deep learning and its clinical applications," Journal of Applied Clinical Medical Physics, 22(1), 11-36, (2021).

[8] S. Pan, J. Flores, C. T. Lin *et al.*, "Generative adversarial networks and radiomics supervision for lung lesion synthesis," SPIE Medical Imaging, (2021).

[9] X. Dong, Y. Lei, S. Tian *et al.*, "Synthetic MRI-aided multi-organ segmentation on male pelvic CT using cycle consistent deep attention network," Radiotherapy and Oncology, 141, (2019).

[10] M. Hu, J. Zhang, L. Matkovic *et al.*, "Reinforcement learning in medical image analysis: Concepts, applications, challenges, and future directions," Journal of Applied Clinical Medical Physics, (2023).

[11] O. Ronneberger, P. Fischer, and T. Brox, "U-Net: Convolutional Networks for Biomedical Image Segmentation," International Conference on Medical Image Computing and Computer-Assisted Intervention, (2015).

[12] H. Lu, H. Wang, Q. Zhang *et al.*, "A 3D Convolutional Neural Network for Volumetric Image Semantic Segmentation," Procedia Manufacturing, 39, 422-428, (2019).

[13] S. Pan, Z. Tian, Y. Lei *et al.*, "CVT-Vnet: a convolutional-transformer model for head and neck multi-organ segmentation," SPIE Medical Imaging, (2022).

[14] A. Hatamizadeh, V. Nath, Y. Tang *et al.*, "Swin UNETR: Swin Transformers for Semantic Segmentation of Brain Tumors in MRI Images," International Workshop, BrainLes 2021, Held in Conjunction with MICCAI 2021, (2021).

[15] A. Hatamizadeh, Y. Tang, V. Nath *et al.*, "UNETR: Transformers for 3D Medical Image Segmentation," IEEE/CVF Winter Conference on Applications of Computer Vision, (2022).

[16] S. Pan, Y. Lei, T. Wang *et al.*, "Male pelvic multi-organ segmentation using token-based transformer Vnet," Physics in Medicine & Biology, 67(20), 205012, (2022).

[17] S. Pan, C.-W. Chang, T. Wang *et al.*, "Abdomen CT multi-organ segmentation using token-based MLP-Mixer," Medical Physics, (2022).

[18] I. Tolstikhin, N. Houlsby, A. Kolesnikov *et al.*, "MLP-Mixer: An all-MLP Architecture for Vision," Conference on Neural Information Processing System, (2021).

[19] J. M. J. Valanarasu, and V. M. Patel, "UNeXt: MLP-based Rapid Medical Image Segmentation Network," International Conference on Medical Image Computing and Computer-Assisted Intervention, (2022).

[20] X. Chen, C. Xie, M. Tan *et al.*, "Robust and Accurate Object Detection via Adversarial Learning," IEEE/CVF Conference on Computer Vision and Pattern Recognition, (2021).

[21] T. Chen, Y. Cheng, Z. Gan *et al.*, "Adversarial Feature Augmentation and Normalization for Visual Recognition," Transactions on Machine Learning Research, (2021).

[22] S.-Y. Lo, P. Oza, and V. M. Patel, "Adversarially Robust One-class Novelty Detection," IEEE Transactions on Pattern Analysis and Machine Intelligence, PP, (2022).

[23] S.-Y. Lo and V. M. Patel, "Defending Against Multiple and Unforeseen Adversarial Videos," IEEE Transactions on Image Processing, 31, 962-973, (2021).

[24] C. Xie, M. Tan, B. Gong *et al.*, "Adversarial Examples Improve Image Recognition," IEEE/CVF Conference on Computer Vision and Pattern Recognition, 816-825, (2020).

[25] I. J. Goodfellow, J. Shlens, and C. Szegedy, "Explaining and Harnessing Adversarial Examples," International Conference on Learning Representations, (2015).

[26] B. Landman, Z. Xu, J. Igelsias *et al.*, "Miccai multi-atlas labeling beyond the cranial vault–workshop and challenge," Proc. MICCAI Multi-Atlas Labeling Beyond Cranial Vault—Workshop Challenge, (2015).
[27] H. Zhang, M. Cisse, Y. N. Dauphin, D. Lopez-Paz, "Mixup: Beyond Empirical Risk Minimization," International Conference on Learning Representations, (2018).
[28] F. Isensee, P. F. Jaeger, S. A. A. Kohl *et al.*, "nnU-Net: a self-configuring method for deep learning-based biomedical image segmentation," Nature Methods, 18(2), 203-211, (2021).